\documentclass[reprint,superscriptaddress,amsmath,amssymb,aps,pre]{revtex4-1}
\usepackage{graphicx}
\usepackage{dcolumn}
\usepackage{bm}
\usepackage{chemist}

\begin{document}


\title{Mechanically-controllable strong 2D ferroelectricity and anisotropic optical properties of flexible BiN monolayer}
\author{Peng Chen}
\affiliation{Beijing National Center for Condensed Matter Physics, Institute of Physics, Chinese Academy of Sciences, Beijing 100190, China.}
\affiliation{School of Physical Sciences, University of Chinese Academy of Sciences, Beijing 100190, China.}
\author{Xue-Jing Zhang}
\affiliation{Beijing National Center for Condensed Matter Physics, Institute of Physics, Chinese Academy of Sciences, Beijing 100190, China.}
\affiliation{School of Physical Sciences, University of Chinese Academy of Sciences, Beijing 100190, China.}
\author{Bang-Gui Liu}
 \email{bgliu@iphy.ac.cn}
 \affiliation{Beijing National Center for Condensed Matter Physics, Institute of Physics, Chinese Academy of Sciences, Beijing 100190, China.}
\affiliation{School of Physical Sciences, University of Chinese Academy of Sciences, Beijing 100190, China.}

\date{\today}

\begin{abstract}
Structural, electronic, ferroelectric, and optical properties of two-dimensional (2D) BiN monolayer material with phosphorene-like structure are studied in terms of the density functional theory and modern Berry phase ferroelectric method. Both phonon spectra and molecular dynamics simulations indicate that the BiN monolayer is a room-temperature stable 2D ferroelectric with polarization as large as 580 pC/m.
Further studies show that the polarization in the BiN monolayer can be easily switched from [100] to [010] direction over the bridging saddle phase by applying a tensile [010] stress of 2.54 N/m or compressive [100] stress of -1.18 N/m. This phase transitions makes its lattice constants vary in a large range compared to other non-ferroelectric 2D materials. Moreover, through applying uniaxial tensile stress parallel to the polarization, one can fix the polarization and change the semiconductor energy gap from direct to indirect one. The optical properties feature a very strong anisotropy in reflectivity below the photon energy of 4 eV.
All these significant ferroelectric, electronic, and optical properties make us believe that the 2D BiN monolayer can be used to make stretchable electronic devices and optical applications.
\end{abstract}

\pacs{Valid PACS appear here}

\maketitle


\section{\label{sec:level1}Introduction}

The two-dimensional material has attracted more and more attentions in both condensed matter physics and material science
due to its wealth of physical phenomena because the charges and spins are confined to a two-dimensional plane\cite{rev_van_der_waals,rev_synthesis,rev_phosphorene,rev_valleytronics,rev_semiconductor,rev_optoelectronics,rev_1}.
These special structural and electronic properties can produce many unique optical, mechanical, and electronic functions\cite{rev_1} and electronic devices\cite{ferro_applications, ferro_photoferroics, ferro_rev}.
In most of the applications of ferroelectric materials, thin films are used, as usual\cite{ferro_thin_film_1, ferro_thin_film_2, ferro_thin_film_3}, because this allows an achievable moderate voltage to switch the polarization\cite{ferro_thin_film}. However, when using thin films, to guarantee the devices work reliably, the quality of the thin film samples and the interfaces are highly demanded. Moreover, there exists a critical thickness\cite{ferro_thickness} for the ferroelectricity in the traditional perovskite ferroelectric ultrathin films, because of the imperfect screening of depolarizing field at the ferroelectric-metal interfaces\cite{ferro_book}. Therefore, a naturally Van der Waals layered ferroelectric materials may be advantageous.

The idea of two-dimensional ferroelectricity is very attractive because 2D layered materials, such as Black phosphorus, graphene, and MoS$_2$, usually have great mechanical flexibility and can sustain large strains $(\sim 25\%)$\cite{lowD_flexible_1,lowD_flexible_2,lowD_flexible_3,Wei_2014}, and ferroelectricity often coexists with tensile strains\cite{ferro_book,ferro_rev,ferro_thin_film}. Unlike the low-dimensional ferromagnetism whose Curie temperature $T_c$ is often far below room temperature (for two-dimensional isotropic Heisenberg spin models, one has $T_c=0$ according to Mermin-Wagner theorem\cite{Mermin_Wagner}), two-dimensional ferroelectricity may survive at a relatively high temperature.
It is also of interest that two-dimensional ferroelectricity is compatible with two-dimensional semiconductors\cite{ferro_book}.
Recently, ferroelectric and even multiferroic phenomenon have been proved in several types of two-dimensional materials\cite{lowD_P1,lowD_P2,lowD_P3,lowD_SnO}. Because the ridged structure of black phosphorus has a unique atomic arrangement order\cite{BP}, ferroelectricity is found in phosphorene-like structures\cite{lowD_P1,lowD_P2,lowD_P3}.

In this paper, we use the Bi 6s lone pair to induce a strong ferroelectric polarization in BiN monolayer with a phosphorene-like structure. Our first-principles investigation proves that this two-dimensional BiN is thermally and dynamically stable. Strain engineering calculations indicate that the ferroelectric polarization and the semiconductor band gap can be easily manipulated by applying uniaxial strains. The mechanical properties prove that the BiN monolayer is  mechanically flexible. Its anisotropic features have a significant effect on its optical properties. These features make us believe that this BiN monolayer with Phosphorene-like structure may be applicable for stretchable electronic devices and optical applications. More detailed results will be presented in the following.

\section{\label{sec:level1}Method}

The calculations are performed in terms of the density functional theory (DFT)\cite{DFT_1,DFT_2} and projector-augmented wave potentials\cite{vasp_paw}, as implemented in VASP\cite{vasp}. We use the exchange-correlation functional PBEsol\cite{PBEsol} because it is best for solid materials.
We adopt usual atomic pseudopotentials of Bi:6s$^2$5d$^10$6p$^3$ and N:2s$^2$2p$^3$. Our computational model is a supercell consisting of the BiN monolayer and a vacuum layer of 20 \AA{}.
The plane wave energy cutoff is set to 500 eV. The $11\times 8\times 1$ k-point mesh in the Brillouin zone is constructed within Monkhorst-Pack scheme\cite{mp}.
Phonopy package\cite{phonopy} is used in the phonon spectrum calculations.
All the structures have been fully relaxed until the largest forces between the atoms become less than 1 meV/\AA{}.
Phase-switching paths were determined with the help of the `Nudged Elastic Band'
(NEB)\cite{neb} method which can give the most energetically favorable intermediate configuration between the initial and final structures.
To calculate the ferroelectric polarization, we use the modern Berry phase method\cite{berry}.
When estimating the polarization of the 3D-stacked BiN structure, the Van der Waals correction was included in the GGA scheme. Ab-initial molecular dynamics simulations are adopted to show the finite temperature stability of ferroelectric phase of the BiN monolayer.

\section{\label{sec:level1}Results}

\subsection{\label{sec:level2}Stable structure and ferroelectricity}

\begin{figure}[!htb]
 \includegraphics*[width=\linewidth]{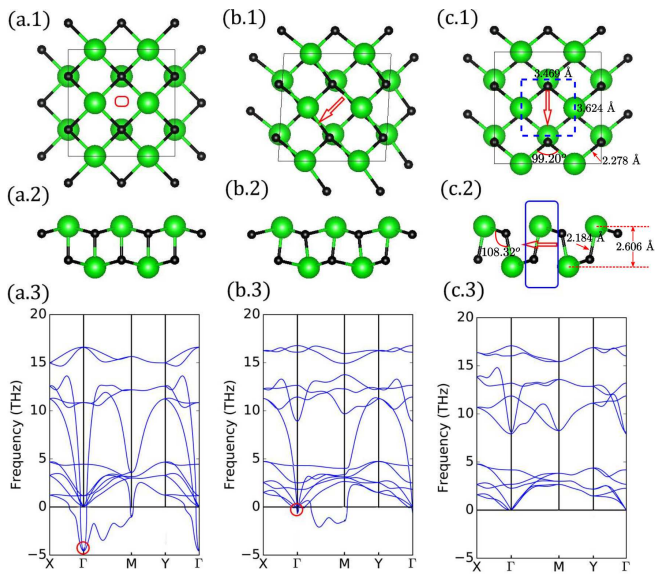}
 \caption{
 \label{fig:structure}
 (Color online) Top views, side views, and phonon spectra of reference phase (a.1, a.2, a.3), saddle phase (b.1, b.2, b.3), and ferroelectric phase (c.1, c.2, c.3) of BiN monolayer. The large green ball represent Bi and the black ball N. The red circle in reference means inversion center, and the red arrows in the structures indicate the ferroelectric distortions.}
\end{figure}

It is well known that the strong ferroelectricity in BiFeO$_3$ originates  from the lone pair of Bi atom. We believe that Bi can cause two-dimensional ferroelectricity in good structures. Considering that the radius of Bi$^{3+}$ ion is almost the same as that of Gd$^{3+}$ ion, we construct the BiN monolayer in terms of the structure of GdN compound. On the other hand, this BiN monolayer can be considered to be made by substituting the P atoms in the famous phosphorene structure by Bi and N atoms. It has center inversion symmetry, as shown in Fig.~\ref{fig:structure} (a.1) and (a.2). It can be seen that the centers of anions and cations coincide, which indicates no ferroelectric polarizations. We will call this phosphorene-like structure as reference phase in the following.

We calculate its zero K phonon spectrum, and find that the largest soft mode frequency can be found at $\Gamma$ point, as shown in Fig.~\ref{fig:structure} (a.3). Further analysis shows that this imaginary frequency is two-fold degenerate and they imply possible ferroelectric displacements along [100] and/or [010] direction.
Then, we distort the reference structure according to the two soft phonon modes.
When applying the two modes together, we can get a ferroelectric phase described in Fig.~\ref{fig:structure} (b.1) and (b.2), and its polarization is along [110] direction. Further calculated phonon spectra show that the structure with [110] polarization still has soft modes, as shown in Fig. 1(b.3), which indicates this phase is a saddle point on the energy surface (saddle phase). When adopting only one of the modes, we obtain the ferroelectric phase in Fig.~\ref{fig:structure} (c.1) and (c.2) and the polarization is along either [100] or [010] direction. The space group of the supercell is Pmn2$_1$ (\#31), and the lattice constants of the supercell are $a=3.469$\AA, $b=3.624$\AA, and $c=15$\AA. The internal coordinates of Bi is (0.5000,0.0000,0.4132) and those of N (0.3866,0.0000,0.5562). There are four atom planes, in the series of Bi-N-N-Bi, and the Bi-Bi plane distance is 2.606\AA. The Bi-N bond lengthes are 2.184\AA{} and 2.278\AA{}.
Further phonon spectra show that there is no imaginary frequency in this case, as shown in Fig. 1(c.3), and therefore the structure with [100] or [010] polarization is dynamically stable.

\begin{figure}[!htb]
 \includegraphics*[width=\linewidth]{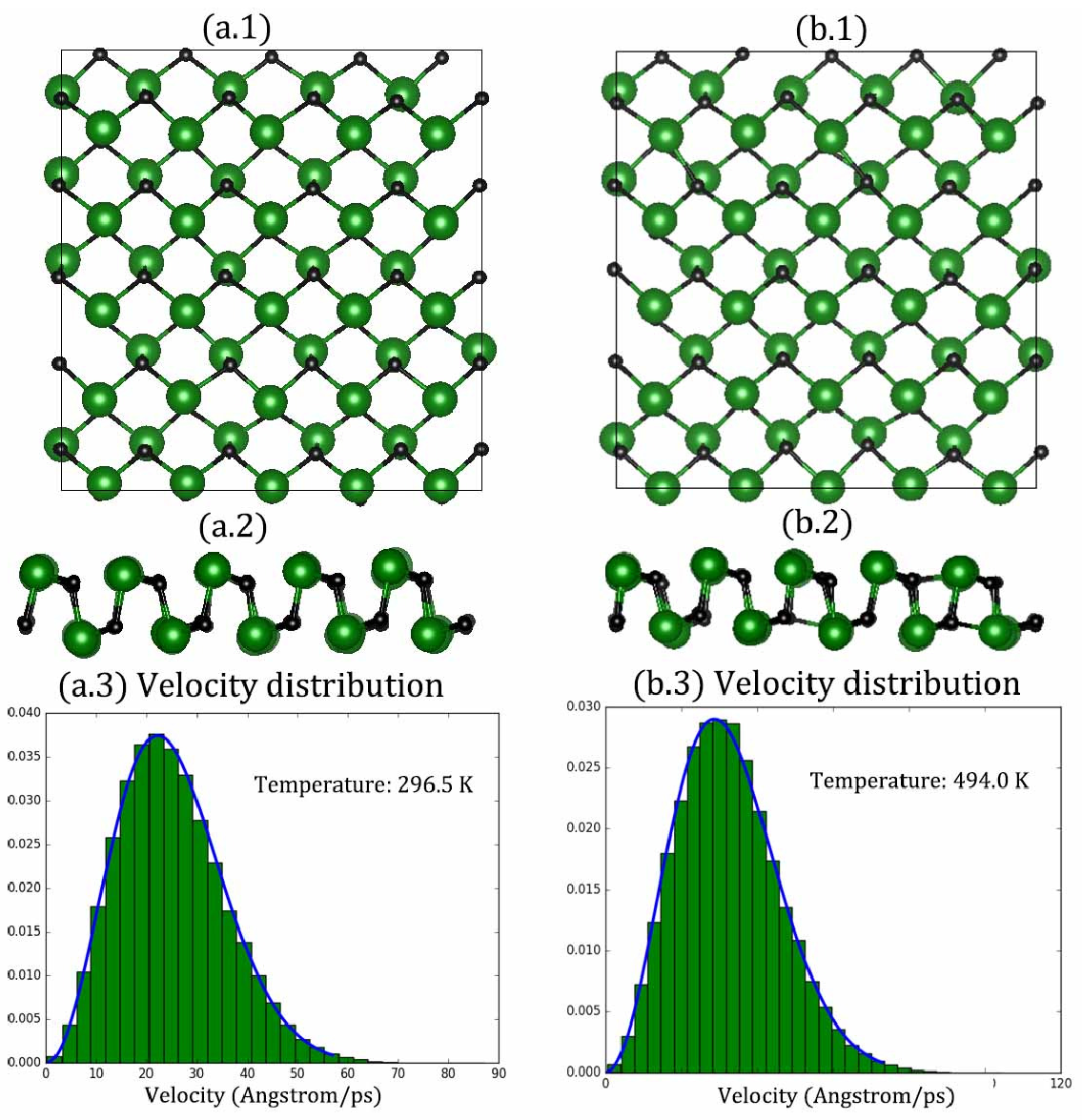}
 \caption{
 \label{fig:Tc}
 (Color online)  Top view snapshots, side view snapshots, and velocity distributions of ferroelectric BiN monolayer at 300 K (a.1, a.2, a.3) and 500 K (b.1, b.2, b.3), from the ab-initio molecular-dynamics simulations for 7.5 ps.}
\end{figure}

Furthermore, we perform ab-initial molecular dynamic simulations of the BiN monolayer at 300 K, 400 K, 500 K, 700 K, and 1100 K. The structures of 300 K and 500K at 7.5 ps are shown in Fig.~\ref{fig:Tc}. We can see that the two-dimensional BiN monolayer can preserve its ferroelectric properties at least up to 500 K; and when the temperature is as high as 700 K, the structure is still stable, but the polarization is substantially reduced; and when the temperature is higher than 1100 K, the structure is broken by the thermal fluctuation.
Therefore, we believe that the ferroelectric two-dimensional BiN monolayer is thermodynamically stable beyond 500 K.

It is noted that the ferroelectric [100] and [010] phases are equivalent to each other, with their polarization orientating in the two directions, but we still distinguish them by their different polarization directions because these equivalent phases can have different behavior under different strains. This is crucial to the discussion of the ferroelectric switching in the following.

\subsection{\label{sec:level2}Mechanical flexibility}

The elastic stiffness constants $C_{ij}$ can be determined by making six finite distortions of the lattice and deriving the elastic constants from the strain-stress relationship\cite{Le_Page_2002}. Since the system is treated as a bulk (in the VASP package) which is a combination of the monolayer and vacuum, the $C_{ij}$ are in GPa. As a result, the 2D elastic stiffness constants should be recovered by $C^{2D}_{ij}=C_{ij}\times c$, in which $c$ is the lattice constant containing the vacuum, accordingly the $C^{2D}_{ij}$ are in J/m$^2$. Then, we can derive Young's modulus ($Y^{2D}$), shear modulus ($G^{2D}$), and Poisson's ratios ($\nu^{2D}$) for the 2D system\cite{Elahi_2015}:
\begin{equation} \label{eq:two-dimensional_Y}
\left\{
\begin{array}{l}
\displaystyle Y^{2D}_{\parallel}=\frac{C^{2D}_{11}C^{2D}_{22}-C^{2D}_{12}C^{2D}_{21}}{C^{2D}_{22}},  \\
\displaystyle Y^{2D}_{\perp}=\frac{C^{2D}_{11}C^{2D}_{22}-C^{2D}_{12}C^{2D}_{21}}{C^{2D}_{11}}, \\
\displaystyle G^{2D}=C^{2D}_{66},\\
\displaystyle \nu^{2D}_{\parallel}=\frac{C^{2D}_{21}}{C^{2D}_{22}}, ~~~
 \nu^{2D}_{\perp}=\frac{C^{2D}_{12}}{C^{2D}_{11}} \end{array}\right.
\end{equation}

\begin{table}[!htb]
\caption{ The two-dimensional Young's modulus (Y$^{2D}_\parallel$ and Y$^{2D}_\perp$ in J/$m^2$), shear modulus (G in J/$m^2$), and Poisson's ratios ($\nu^{2D}_{\parallel}$ and $\nu^{2D}_{\perp}$) for the Phosphorene, Graphene, \chemform{MoS_2}, and monolayer BiN. The $\parallel$ and $\perp$ indicate the ferroelectric direction and the direction perpendicular to the ferroelectric orientation respectively.}
\label{tab:modulus}
\begin{ruledtabular}
\begin{tabular}{lcccc}
       & Phosphorene\cite{Wei_2014} & Graphene\cite{Bosak_2007} & \chemform{MoS_2} \cite{Liu_2014} & BiN\\
\hline
$Y^{2D}_\parallel$    & 24.42 & 366.4 & 123  & 47.14\\
$Y^{2D}_\perp$       & 92.13 & 366.4 & 123  & 101.41\\
$G^{2D}$                  & 22.75 & 162.8 & 47.9\cite{Peng_2013} & 11.78\\
$\nu^{2D}_{\parallel}$              & 0.17  & 0.125 & 0.25 & 0.08\\
$\nu^{2D}_{\perp}$                  & 0.62  & 0.125 & 0.25 & 0.21\\
\end{tabular}
\end{ruledtabular}
\end{table}

Our calculated Young's modulus and Poisson's ratios are listed in Table~\ref{tab:modulus}. Due to the polarization, the Young's modulus (Poisson's ratio) are anisotropic in the plane of the monolayer, being different when the direction is parallel ($\parallel$) and perpendicular ($\perp$) to the ferroelectric orientation. The Young's modulus in the $\perp$ direction is two times larger than that in the $\parallel$ direction.
As shown in Table~\ref{tab:modulus}, Graphene has the largest Young's and shear modulus, and MoS$_2$ and phosphorene have smaller values. It can be seen that the moduli of the BiN monolayer are larger than those of phosphorene and the in-plane anisotropy is less than that of phosphorene. Although the moduli are the second smallest among the four monolayer systems, the shear modulus of the BiN monolayer is the smallest, nearly half that of phosphorene.
For the Poisson's ratio, we notice that the BiN monolayer has the smallest $\nu^{2D}_{\parallel}$, which means that when we squeeze (stretch) the BiN monolayer along the ferroelectric direction, it expands (shrinks) the smallest in the perpendicular direction. We believe that the BiN monolayer can preserve its superior mechanical flexibility as phosphorene does.

\subsection{\label{sec:level2}Ferroelectric polarization and mechanical switching}

\begin{figure}[!htb]
 \includegraphics*[width=\linewidth]{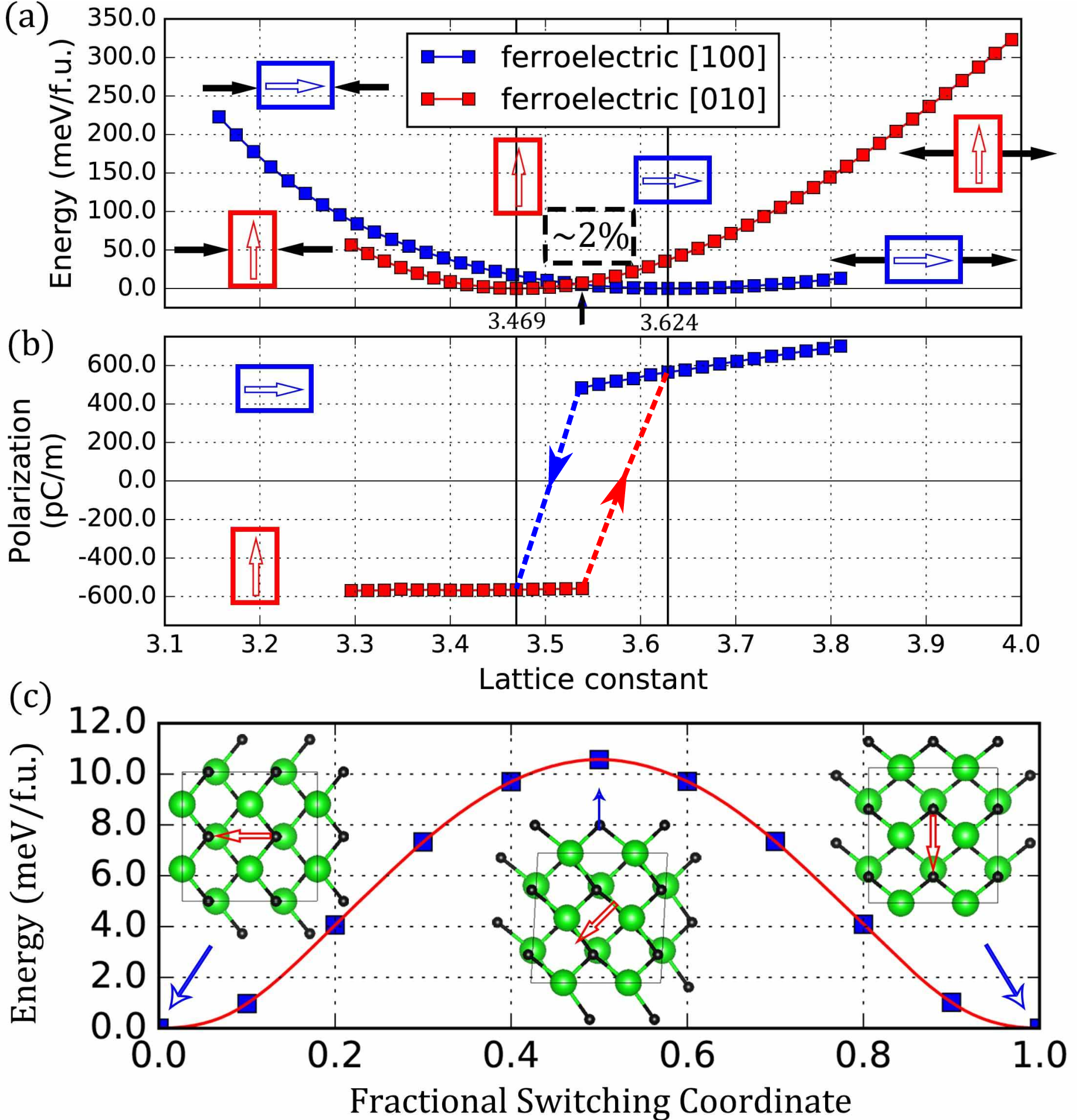}
 \caption{
 \label{fig:strain_neb}
 (Color online) Total energies (a) and ferroelectric polarizations (b) of ferroelectric [100] and [010] phases as functions of the lattice constant. The two black dashed vertical lines indicate the equivalent lattice constants in the two directions. (c) Energy values along the switching path from ferroelectric [010] phase to ferroelectric [100] phase through the saddle phase whose polarization is along [110] direction.}
\end{figure}

We calculate the ferroelectric polarization with the modern berry phase ferroelectric theory. The polarization of the BiN monolayer reaches to 580 pC/m which is larger than that of the predicted GeSn. If we stack the BiN monolayers to form a 3D layered structure, the polarization can be as large as 88.8 $\mu$C/cm$^2$ which is comparable with the famous room temperature multiferroic BiFeO$_3$\cite{BFO_1,BFO_2,BFO_3}.
The ferroelectric polarization of the BiN monolayer can take one of the [100] and [010] directions, and then we think that the BiN monolayer can take one of the two phases defined to have polarization along the [100] and [010] directions, respectively. The uniaxial stresses parallel and perpendicular to the polarization will produce different effects on the BiN monolayer. When a uniaxial stress is applied, there will be a corresponding strain parallel to the stress and an opposite strain perpendicular to the stress. The perpendicular strain is determined by optimizing the total energy along this direction. In Fig.~\ref{fig:strain_neb}~(a), we present the total energies of the two ferroelectric phases as functions of the parallel strains. The two vertical black lines indicate the equivalent lattice constants along [100] direction (lattice constant $a$) for the [100] and [010] phases. If focusing on the red line (ferroelectric [010]) in Fig.~\ref{fig:strain_neb}~(a), the strains at the left lattice constant means the compressive strains perpendicular to the polarization direction and the strains at the right lattice constant indicate the tensile strain perpendicular to the polarization direction. The same configurations are applied to the blue line (ferroelectric [100]) in Fig.~\ref{fig:strain_neb}~(a), the only difference is that the strains are parallel to the polarization direction. We can see that there is an energy crossing between the ferroelectric [100] (blue line in Fig.~\ref{fig:strain_neb}) and the ferroelectric [010] (red line in Fig.~\ref{fig:strain_neb}) at $a = 3.5388$\AA, which means that there will be a phase transition at this point, and accordingly a polarization switching between the [100] and [010] directions is shown in Fig.~\ref{fig:strain_neb}~(b).

The strain  to induce the phase transition of the the [100] phase is only $2\%$. This strain can be produced by applying a parallel compressive stress of -1.18 N/m, or a perpendicular tensile stress of 2.54 N/m,  which can be easily achieved in the two-dimensional materials.
The energy barrier between these two phases has been calculated with the NEB method, as shown in Fig.~\ref{fig:strain_neb}~(c). We find out that the energy barrier is equivalent to 81 meV/f.u. if the ferroelectric [100] phase is switched through the reference phase (Fig.~\ref{fig:structure} (a.1) and (a.2)), but the energy barrier reduces to 10.5 meV/f.u. if the ferroelectric [100] phase is switched through the saddle phase (Fig.~\ref{fig:structure} (b.1) and (b.2)) that has polarization along the [110] direction.
Because of the stress-induced phase transition, the lattice constant of the BiN monolayer can vary in a very large range, which makes it very flexible.
From the blue line of the tensile strain part (ferroelectric [100]) in Fig.~\ref{fig:strain_neb}~(a), we can also see that the energy increases very slowly with the strain, which indicates the BiN monolayer is very mechanically flexible. On the other hand, we can also fix the polarization by applying tensile stress parallel to the polarization, as shown in Fig. 3(a).
All these properties can make the BiN monolayer a very promising candidate for stretchable electronic devices.

\subsection{\label{sec:level2} Mechanical manipulation of optical properties}

\begin{figure}[!htb]
 \includegraphics*[width=\linewidth]{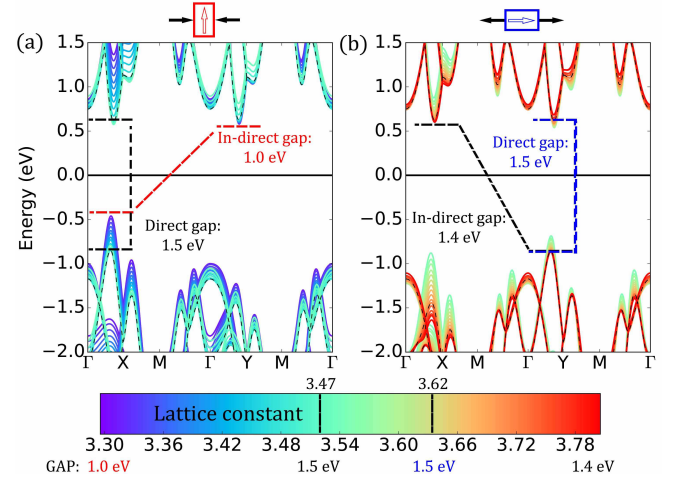}
 \caption{
 \label{fig:Ebands}
 (Color online) Energy bands of the the ferroelectric [010] (a) and [100] (b) phases, with the [100] lattice constant changing from 3.30 to 3.80 \AA{}.}
\end{figure}

The band structures of the BiN monolayer under a series of uniaxial stresses are calculated. We have found that the energy band gaps can be tuned by the uniaxial stress, changing from direct gap to indirect gap. In Fig.~\ref{fig:Ebands}, the black dashed lines in (a) and (b) indicate the band structures of the ferroelectric [010] and [100] phases, respectively. In the Fig.~\ref{fig:Ebands} (a), we can see that it is a direct gap of 1.5 eV for the ferroelectric [010] phase, and it becomes an indirect band gap when compressive uniaxial stress is applied, reaching to an indirect gap of 1.0 eV at the parallel strain of $-5\%$. In the Fig.~\ref{fig:Ebands} (b), the direct gap is 1.5 eV at the equilibrium position, and the indirect gap is about 1.4 eV for a parallel tensile strain of $5\%$.
Comparing (a) and (b) in Fig.~\ref{fig:Ebands}, we can see that the gap transition, from direct to indirect, is much easy to achieve by applying uniaxial stress perpendicular to the polarization direction, but it cannot be tuned much by applying uniaxial stress parallel to the polarization direction. Therefore, while stretching the BiN momolayer as a two-dimensional material, we can achieve not only directional switching of its ferroelectric polarization but also manipulation of its semiconductor band gap.

\begin{figure}[!htb]
 \includegraphics*[width=0.9\linewidth]{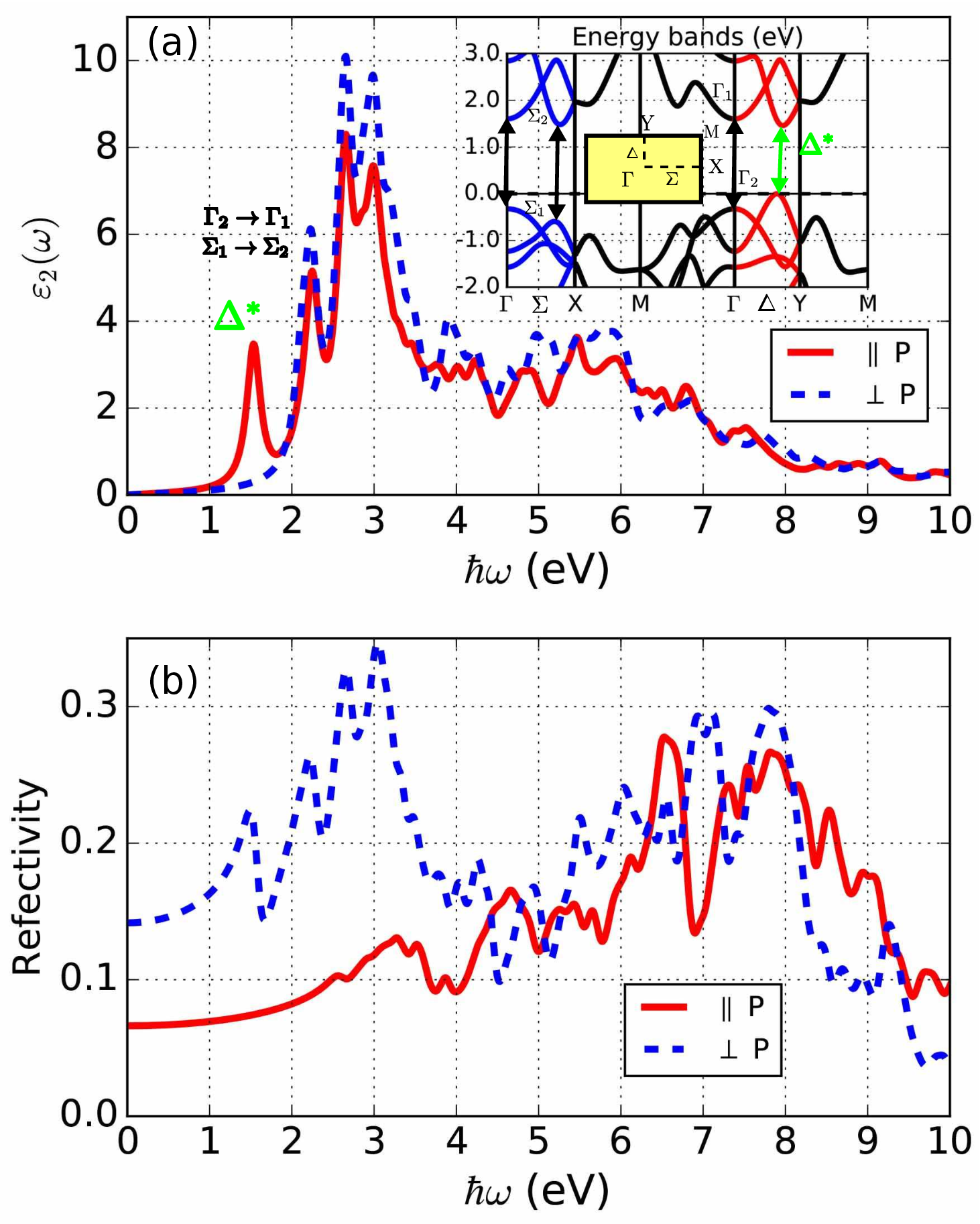}
 \caption{ \label{fig:optics}
 (Color online) The imaginary part of dielectric function (a) and reflectivity (b) of the BiN monolayer as functions of photon energy parallel (red) and perpendicular (blue) to the ferroelectric directions. The inset describes the partial energy band structure with arrows indicating the transitions.}
\end{figure}

We calculate frequency-dependent dielectric constants  and present the dielectric functions parallel and perpendicular to the ferroelectric directions in Fig.~\ref{fig:optics}. As shown in the anisotropic Young's modulus, the anisotropic feature is also reflected in the optical properties of the BiN monolayer. The peak $\Delta^{\star}$ along the ferroelectric direction can only be found around 1.5 eV in the Fig. ~\ref{fig:optics}(a). This unique peak comes from the direct band gap along the $\Gamma\rightarrow X$ ($\Delta$) direction. Since the compressive stress applied to the $\perp$ direction can tune the direct gap into an indirect gap, the $\Delta^{\star}$ feature can be changed by applying such stress. The reflectivity below 4 eV is strongly anisotropic, as shown in Fig. ~\ref{fig:optics}(b), which implies that for the photon energy between 0 and 3.5 eV, the perpendicular reflectivity is at least twice the parallel reflectivity. Such strong anisotropy should be useful for designing new applications.

\section{\label{sec:level1}Discussion and Conclusion}

The crystal structure of the BiN monolayer originates from the famous faced center cubic GdN\cite{GdN_1}.
It is well known that GdN can preserve ferroelectric under strains\cite{GdN_2}. It is reasonable to believe that the ferroelectric distortion without strain in the BiN monolayer comes from the Bi$^{3+}$ lone-pair 6s electrons. The BiN monolayer is mechanically flexible according to our Young's modulus calculation, being comparable with the phosphorene which has been proved to be the superior mechanical flexibility material\cite{Wei_2014}. The shear modulus of the BiN monolayer is even smaller than that of phosphorene, which explains why the energy barrier is so small for switching the ferroelectric polarization from [100] to [010] direction through the saddle phase. From the Fig.~\ref{fig:strain_neb}, the ferroelectric phase transition between [100] and [010] can be induced by 2$\%$ strain, corresponding to the lattice constant ranging from 3.469 \AA{} to 3.642 \AA{}. Considering the superior mechanical flexibilities, the BiN monolayer can be very easily stretched in a large range of lattice constant. Moreover, due to the ferroelectric polarization, the symmetry was broken between the [100] and [010] direction, which induces the anisotropic features of the energy band structure and the optical properties.\\

In summary, we have studied a two-dimensional BiN monolayer with phosphorene-like structure which has been proven to tending to preserve ferroelectricity. Our calculated phonon spectra and MD calculations have show its structural stability. Our DFT and Berry phase based modern ferroelectric theory studies have shown that the BiN monolayer has a 2D ferroelectricity with as large polarization as 580 pC/m, being comparable with BiFeO$_3$ in the 3D case. Further mechanical studies show that the polarization in this BiN monolayer can be easily switched from [100] to [010] direction over a saddle phase  through overcoming a 10.5 meV/f.u. barrier by applying small tensile [010] stress of 2.54 N/m or compressive [100] stress of -1.18 N/m, corresponding to a strain of 2\%. This phase transition in this 2D BiN monolayer makes its lattice constant change in a very large range in comparison with other non-ferroelectric materials. Moreover, through applying uniaxial tensile stress perpendicular to the polarization, one can fix the ferroelectric polarization and change the semiconductor energy gap, between direct and indirect. A very strong anisotropy has been found in the optical reflectivity when the photon energy is below 4 eV. All these features make us believe that the BiN monolayer as a two-dimensional material can be used to achieve stretchable electronic devices and optical applications.

\begin{acknowledgments}
This work is supported by the Nature Science Foundation of China (No.11574366), by the Strategic Priority Research Program of the Chinese Academy of Sciences (Grant No.XDB07000000), and by the Department of Science and Technology of China (Grant No.2016YFA0300701). The calculations were performed in the Milky Way \#2 supercomputer system at the National Supercomputer Center of Guangzhou, Guangzhou, China.
\end{acknowledgments}

\providecommand{\noopsort}[1]{}\providecommand{\singleletter}[1]{#1}%


\begin{thebibliography}{45}%
\makeatletter
\providecommand \@ifxundefined [1]{%
 \@ifx{#1\undefined}
}%
\providecommand \@ifnum [1]{%
 \ifnum #1\expandafter \@firstoftwo
 \else \expandafter \@secondoftwo
 \fi
}%
\providecommand \@ifx [1]{%
 \ifx #1\expandafter \@firstoftwo
 \else \expandafter \@secondoftwo
 \fi
}%
\providecommand \natexlab [1]{#1}%
\providecommand \enquote  [1]{``#1''}%
\providecommand \bibnamefont  [1]{#1}%
\providecommand \bibfnamefont [1]{#1}%
\providecommand \citenamefont [1]{#1}%
\providecommand \href@noop [0]{\@secondoftwo}%
\providecommand \href [0]{\begingroup \@sanitize@url \@href}%
\providecommand \@href[1]{\@@startlink{#1}\@@href}%
\providecommand \@@href[1]{\endgroup#1\@@endlink}%
\providecommand \@sanitize@url [0]{\catcode `\\12\catcode `\$12\catcode
  `\&12\catcode `\#12\catcode `\^12\catcode `\_12\catcode `\%12\relax}%
\providecommand \@@startlink[1]{}%
\providecommand \@@endlink[0]{}%
\providecommand \url  [0]{\begingroup\@sanitize@url \@url }%
\providecommand \@url [1]{\endgroup\@href {#1}{\urlprefix }}%
\providecommand \urlprefix  [0]{URL }%
\providecommand \Eprint [0]{\href }%
\providecommand \doibase [0]{http://dx.doi.org/}%
\providecommand \selectlanguage [0]{\@gobble}%
\providecommand \bibinfo  [0]{\@secondoftwo}%
\providecommand \bibfield  [0]{\@secondoftwo}%
\providecommand \translation [1]{[#1]}%
\providecommand \BibitemOpen [0]{}%
\providecommand \bibitemStop [0]{}%
\providecommand \bibitemNoStop [0]{.\EOS\space}%
\providecommand \EOS [0]{\spacefactor3000\relax}%
\providecommand \BibitemShut  [1]{\csname bibitem#1\endcsname}%
\let\auto@bib@innerbib\@empty
\bibitem [{\citenamefont {Novoselov}\ \emph {et~al.}(2016)\citenamefont
  {Novoselov}, \citenamefont {Mishchenko}, \citenamefont {Carvalho},\ and\
  \citenamefont {Neto}}]{rev_van_der_waals}%
  \BibitemOpen
  \bibfield  {author} {\bibinfo {author} {\bibfnamefont {K.~S.}\ \bibnamefont
  {Novoselov}}, \bibinfo {author} {\bibfnamefont {A.}~\bibnamefont
  {Mishchenko}}, \bibinfo {author} {\bibfnamefont {A.}~\bibnamefont
  {Carvalho}}, \ and\ \bibinfo {author} {\bibfnamefont {A.~H.~C.}\ \bibnamefont
  {Neto}},\ }\href@noop {} {\bibfield  {journal} {\bibinfo  {journal}
  {Science}\ }\textbf {\bibinfo {volume} {353}},\ \bibinfo {pages} {9439}
  (\bibinfo {year} {2016})}\BibitemShut {NoStop}%
\bibitem [{\citenamefont {Mannix}\ \emph {et~al.}(2017)\citenamefont {Mannix},
  \citenamefont {Kiraly}, \citenamefont {Hersam},\ and\ \citenamefont
  {Guisinger}}]{rev_synthesis}%
  \BibitemOpen
  \bibfield  {author} {\bibinfo {author} {\bibfnamefont {A.~J.}\ \bibnamefont
  {Mannix}}, \bibinfo {author} {\bibfnamefont {B.}~\bibnamefont {Kiraly}},
  \bibinfo {author} {\bibfnamefont {M.~C.}\ \bibnamefont {Hersam}}, \ and\
  \bibinfo {author} {\bibfnamefont {N.~P.}\ \bibnamefont {Guisinger}},\
  }\href@noop {} {\bibfield  {journal} {\bibinfo  {journal} {Nature Reviews
  Chemistry}\ }\textbf {\bibinfo {volume} {1}},\ \bibinfo {pages} {0014}
  (\bibinfo {year} {2017})}\BibitemShut {NoStop}%
\bibitem [{\citenamefont {Carvalho}\ \emph {et~al.}(2016)\citenamefont
  {Carvalho}, \citenamefont {Wang}, \citenamefont {Zhu}, \citenamefont {Rodin},
  \citenamefont {Su},\ and\ \citenamefont {Neto}}]{rev_phosphorene}%
  \BibitemOpen
  \bibfield  {author} {\bibinfo {author} {\bibfnamefont {A.}~\bibnamefont
  {Carvalho}}, \bibinfo {author} {\bibfnamefont {M.}~\bibnamefont {Wang}},
  \bibinfo {author} {\bibfnamefont {X.}~\bibnamefont {Zhu}}, \bibinfo {author}
  {\bibfnamefont {A.~S.}\ \bibnamefont {Rodin}}, \bibinfo {author}
  {\bibfnamefont {H.}~\bibnamefont {Su}}, \ and\ \bibinfo {author}
  {\bibfnamefont {A.~H.~C.}\ \bibnamefont {Neto}},\ }\href@noop {} {\bibfield
  {journal} {\bibinfo  {journal} {Nature Reviews Materials}\ }\textbf {\bibinfo
  {volume} {1}},\ \bibinfo {pages} {16061} (\bibinfo {year}
  {2016})}\BibitemShut {NoStop}%
\bibitem [{\citenamefont {Schaibley}\ \emph {et~al.}(2016)\citenamefont
  {Schaibley}, \citenamefont {Yu}, \citenamefont {Clark}, \citenamefont
  {Rivera}, \citenamefont {Ross}, \citenamefont {Seyler}, \citenamefont {Yao},\
  and\ \citenamefont {Xu}}]{rev_valleytronics}%
  \BibitemOpen
  \bibfield  {author} {\bibinfo {author} {\bibfnamefont {J.~R.}\ \bibnamefont
  {Schaibley}}, \bibinfo {author} {\bibfnamefont {H.}~\bibnamefont {Yu}},
  \bibinfo {author} {\bibfnamefont {G.}~\bibnamefont {Clark}}, \bibinfo
  {author} {\bibfnamefont {P.}~\bibnamefont {Rivera}}, \bibinfo {author}
  {\bibfnamefont {J.~S.}\ \bibnamefont {Ross}}, \bibinfo {author}
  {\bibfnamefont {K.~L.}\ \bibnamefont {Seyler}}, \bibinfo {author}
  {\bibfnamefont {W.}~\bibnamefont {Yao}}, \ and\ \bibinfo {author}
  {\bibfnamefont {X.}~\bibnamefont {Xu}},\ }\href@noop {} {\bibfield  {journal}
  {\bibinfo  {journal} {Nature Reviews Materials}\ }\textbf {\bibinfo {volume}
  {1}},\ \bibinfo {pages} {16055} (\bibinfo {year} {2016})}\BibitemShut
  {NoStop}%
\bibitem [{\citenamefont {Chhowalla}\ \emph {et~al.}(2016)\citenamefont
  {Chhowalla}, \citenamefont {Jena},\ and\ \citenamefont
  {Zhang}}]{rev_semiconductor}%
  \BibitemOpen
  \bibfield  {author} {\bibinfo {author} {\bibfnamefont {M.}~\bibnamefont
  {Chhowalla}}, \bibinfo {author} {\bibfnamefont {D.}~\bibnamefont {Jena}}, \
  and\ \bibinfo {author} {\bibfnamefont {H.}~\bibnamefont {Zhang}},\
  }\href@noop {} {\bibfield  {journal} {\bibinfo  {journal} {Nature Reviews
  Materials}\ }\textbf {\bibinfo {volume} {1}},\ \bibinfo {pages} {16052}
  (\bibinfo {year} {2016})}\BibitemShut {NoStop}%
\bibitem [{\citenamefont {Wang}\ \emph {et~al.}(2012)\citenamefont {Wang},
  \citenamefont {Kalantar-Zadeh}, \citenamefont {Kis}, \citenamefont
  {Coleman},\ and\ \citenamefont {Strano}}]{rev_optoelectronics}%
  \BibitemOpen
  \bibfield  {author} {\bibinfo {author} {\bibfnamefont {Q.~H.}\ \bibnamefont
  {Wang}}, \bibinfo {author} {\bibfnamefont {K.}~\bibnamefont
  {Kalantar-Zadeh}}, \bibinfo {author} {\bibfnamefont {A.}~\bibnamefont {Kis}},
  \bibinfo {author} {\bibfnamefont {J.~N.}\ \bibnamefont {Coleman}}, \ and\
  \bibinfo {author} {\bibfnamefont {M.~S.}\ \bibnamefont {Strano}},\
  }\href@noop {} {\bibfield  {journal} {\bibinfo  {journal} {Nature
  Nanotechnology}\ }\textbf {\bibinfo {volume} {7}},\ \bibinfo {pages} {699}
  (\bibinfo {year} {2012})}\BibitemShut {NoStop}%
\bibitem [{\citenamefont {Das}\ \emph {et~al.}(2015)\citenamefont {Das},
  \citenamefont {Robinson}, \citenamefont {Dubey}, \citenamefont {Terrones},\
  and\ \citenamefont {Terrones}}]{rev_1}%
  \BibitemOpen
  \bibfield  {author} {\bibinfo {author} {\bibfnamefont {S.}~\bibnamefont
  {Das}}, \bibinfo {author} {\bibfnamefont {J.~A.}\ \bibnamefont {Robinson}},
  \bibinfo {author} {\bibfnamefont {M.}~\bibnamefont {Dubey}}, \bibinfo
  {author} {\bibfnamefont {H.}~\bibnamefont {Terrones}}, \ and\ \bibinfo
  {author} {\bibfnamefont {M.}~\bibnamefont {Terrones}},\ }\href@noop {}
  {\bibfield  {journal} {\bibinfo  {journal} {Annual Review of Materials
  Research}\ }\textbf {\bibinfo {volume} {45}},\ \bibinfo {pages} {1} (\bibinfo
  {year} {2015})}\BibitemShut {NoStop}%
\bibitem [{\citenamefont {Abrahams}(1993)}]{ferro_applications}%
  \BibitemOpen
  \bibfield  {author} {\bibinfo {author} {\bibfnamefont {S.~C.}\ \bibnamefont
  {Abrahams}},\ }\href@noop {} {\bibfield  {journal} {\bibinfo  {journal}
  {Ferroelectrics}\ }\textbf {\bibinfo {volume} {138}},\ \bibinfo {pages} {307}
  (\bibinfo {year} {1993})}\BibitemShut {NoStop}%
\bibitem [{\citenamefont {Butler}\ \emph {et~al.}(2015)\citenamefont {Butler},
  \citenamefont {Frost},\ and\ \citenamefont {Walsh}}]{ferro_photoferroics}%
  \BibitemOpen
  \bibfield  {author} {\bibinfo {author} {\bibfnamefont {K.~T.}\ \bibnamefont
  {Butler}}, \bibinfo {author} {\bibfnamefont {J.~M.}\ \bibnamefont {Frost}}, \
  and\ \bibinfo {author} {\bibfnamefont {A.}~\bibnamefont {Walsh}},\
  }\href@noop {} {\bibfield  {journal} {\bibinfo  {journal} {Energy \&
  Environmental Science}\ }\textbf {\bibinfo {volume} {8}},\ \bibinfo {pages}
  {838} (\bibinfo {year} {2015})}\BibitemShut {NoStop}%
\bibitem [{\citenamefont {Ramesh}\ and\ \citenamefont
  {Spaldin}(2007)}]{ferro_rev}%
  \BibitemOpen
  \bibfield  {author} {\bibinfo {author} {\bibfnamefont {R.}~\bibnamefont
  {Ramesh}}\ and\ \bibinfo {author} {\bibfnamefont {N.~A.}\ \bibnamefont
  {Spaldin}},\ }\href@noop {} {\bibfield  {journal} {\bibinfo  {journal}
  {Nature Materials}\ }\textbf {\bibinfo {volume} {6}},\ \bibinfo {pages} {21}
  (\bibinfo {year} {2007})}\BibitemShut {NoStop}%
\bibitem [{\citenamefont {Martin}\ and\ \citenamefont
  {Rappe}(2016)}]{ferro_thin_film_1}%
  \BibitemOpen
  \bibfield  {author} {\bibinfo {author} {\bibfnamefont {L.~W.}\ \bibnamefont
  {Martin}}\ and\ \bibinfo {author} {\bibfnamefont {A.~M.}\ \bibnamefont
  {Rappe}},\ }\href@noop {} {\bibfield  {journal} {\bibinfo  {journal} {Nature
  Reviews Materials}\ }\textbf {\bibinfo {volume} {2}},\ \bibinfo {pages}
  {16087} (\bibinfo {year} {2016})}\BibitemShut {NoStop}%
\bibitem [{\citenamefont {Setter}\ \emph {et~al.}(2006)\citenamefont {Setter},
  \citenamefont {Damjanovic}, \citenamefont {Eng}, \citenamefont {Fox},
  \citenamefont {Gevorgian}, \citenamefont {Hong}, \citenamefont {Kingon},
  \citenamefont {Kohlstedt}, \citenamefont {Park}, \citenamefont {Stephenson},
  \citenamefont {Stolitchnov}, \citenamefont {Taganstev}, \citenamefont
  {Taylor}, \citenamefont {Yamada},\ and\ \citenamefont
  {Streiffer}}]{ferro_thin_film_2}%
  \BibitemOpen
  \bibfield  {author} {\bibinfo {author} {\bibfnamefont {N.}~\bibnamefont
  {Setter}}, \bibinfo {author} {\bibfnamefont {D.}~\bibnamefont {Damjanovic}},
  \bibinfo {author} {\bibfnamefont {L.}~\bibnamefont {Eng}}, \bibinfo {author}
  {\bibfnamefont {G.}~\bibnamefont {Fox}}, \bibinfo {author} {\bibfnamefont
  {S.}~\bibnamefont {Gevorgian}}, \bibinfo {author} {\bibfnamefont
  {S.}~\bibnamefont {Hong}}, \bibinfo {author} {\bibfnamefont {A.}~\bibnamefont
  {Kingon}}, \bibinfo {author} {\bibfnamefont {H.}~\bibnamefont {Kohlstedt}},
  \bibinfo {author} {\bibfnamefont {N.~Y.}\ \bibnamefont {Park}}, \bibinfo
  {author} {\bibfnamefont {G.~B.}\ \bibnamefont {Stephenson}}, \bibinfo
  {author} {\bibfnamefont {I.}~\bibnamefont {Stolitchnov}}, \bibinfo {author}
  {\bibfnamefont {A.~K.}\ \bibnamefont {Taganstev}}, \bibinfo {author}
  {\bibfnamefont {D.~V.}\ \bibnamefont {Taylor}}, \bibinfo {author}
  {\bibfnamefont {T.}~\bibnamefont {Yamada}}, \ and\ \bibinfo {author}
  {\bibfnamefont {S.}~\bibnamefont {Streiffer}},\ }\href@noop {} {\bibfield
  {journal} {\bibinfo  {journal} {Journal of Applied Physics}\ }\textbf
  {\bibinfo {volume} {100}},\ \bibinfo {pages} {051606} (\bibinfo {year}
  {2006})}\BibitemShut {NoStop}%
\bibitem [{\citenamefont {Schlom}\ \emph {et~al.}(2007)\citenamefont {Schlom},
  \citenamefont {Chen}, \citenamefont {Eom}, \citenamefont {Rabe},
  \citenamefont {Streiffer},\ and\ \citenamefont
  {Triscone}}]{ferro_thin_film_3}%
  \BibitemOpen
  \bibfield  {author} {\bibinfo {author} {\bibfnamefont {D.~G.}\ \bibnamefont
  {Schlom}}, \bibinfo {author} {\bibfnamefont {L.-Q.}\ \bibnamefont {Chen}},
  \bibinfo {author} {\bibfnamefont {C.-B.}\ \bibnamefont {Eom}}, \bibinfo
  {author} {\bibfnamefont {K.~M.}\ \bibnamefont {Rabe}}, \bibinfo {author}
  {\bibfnamefont {S.~K.}\ \bibnamefont {Streiffer}}, \ and\ \bibinfo {author}
  {\bibfnamefont {J.-M.}\ \bibnamefont {Triscone}},\ }\href@noop {} {\bibfield
  {journal} {\bibinfo  {journal} {Annual Review of Materials Research}\
  }\textbf {\bibinfo {volume} {37}},\ \bibinfo {pages} {589} (\bibinfo {year}
  {2007})}\BibitemShut {NoStop}%
\bibitem [{\citenamefont {Dawber}\ \emph {et~al.}(2005)\citenamefont {Dawber},
  \citenamefont {Rabe},\ and\ \citenamefont {Scott}}]{ferro_thin_film}%
  \BibitemOpen
  \bibfield  {author} {\bibinfo {author} {\bibfnamefont {M.}~\bibnamefont
  {Dawber}}, \bibinfo {author} {\bibfnamefont {K.~M.}\ \bibnamefont {Rabe}}, \
  and\ \bibinfo {author} {\bibfnamefont {J.~F.}\ \bibnamefont {Scott}},\
  }\href@noop {} {\bibfield  {journal} {\bibinfo  {journal} {Reviews of Modern
  Physics}\ }\textbf {\bibinfo {volume} {77}},\ \bibinfo {pages} {1083}
  (\bibinfo {year} {2005})}\BibitemShut {NoStop}%
\bibitem [{\citenamefont {Junquera}\ and\ \citenamefont
  {Ghosez}(2003)}]{ferro_thickness}%
  \BibitemOpen
  \bibfield  {author} {\bibinfo {author} {\bibfnamefont {J.}~\bibnamefont
  {Junquera}}\ and\ \bibinfo {author} {\bibfnamefont {P.}~\bibnamefont
  {Ghosez}},\ }\href@noop {} {\bibfield  {journal} {\bibinfo  {journal}
  {Nature}\ }\textbf {\bibinfo {volume} {422}},\ \bibinfo {pages} {506}
  (\bibinfo {year} {2003})}\BibitemShut {NoStop}%
\bibitem [{\citenamefont {Rabe}\ \emph {et~al.}(2007)\citenamefont {Rabe},
  \citenamefont {Ahn},\ and\ \citenamefont {Triscone}}]{ferro_book}%
  \BibitemOpen
  \bibfield  {author} {\bibinfo {author} {\bibfnamefont {K.~M.}\ \bibnamefont
  {Rabe}}, \bibinfo {author} {\bibfnamefont {C.~H.}\ \bibnamefont {Ahn}}, \
  and\ \bibinfo {author} {\bibfnamefont {J.-M.}\ \bibnamefont {Triscone}},\
  }\href@noop {} {\emph {\bibinfo {title} {Physics of ferroelectrics: a modern
  perspective}}},\ Vol.\ \bibinfo {volume} {105}\ (\bibinfo  {publisher}
  {Springer science \& business media},\ \bibinfo {year} {2007})\BibitemShut
  {NoStop}%
\bibitem [{\citenamefont {Kim}\ \emph {et~al.}(2009)\citenamefont {Kim},
  \citenamefont {Zhao}, \citenamefont {Jang}, \citenamefont {Lee},
  \citenamefont {Kim}, \citenamefont {Kim}, \citenamefont {Ahn}, \citenamefont
  {Kim}, \citenamefont {Choi},\ and\ \citenamefont {Hong}}]{lowD_flexible_1}%
  \BibitemOpen
  \bibfield  {author} {\bibinfo {author} {\bibfnamefont {K.~S.}\ \bibnamefont
  {Kim}}, \bibinfo {author} {\bibfnamefont {Y.}~\bibnamefont {Zhao}}, \bibinfo
  {author} {\bibfnamefont {H.}~\bibnamefont {Jang}}, \bibinfo {author}
  {\bibfnamefont {S.~Y.}\ \bibnamefont {Lee}}, \bibinfo {author} {\bibfnamefont
  {J.~M.}\ \bibnamefont {Kim}}, \bibinfo {author} {\bibfnamefont {K.~S.}\
  \bibnamefont {Kim}}, \bibinfo {author} {\bibfnamefont {J.-H.}\ \bibnamefont
  {Ahn}}, \bibinfo {author} {\bibfnamefont {P.}~\bibnamefont {Kim}}, \bibinfo
  {author} {\bibfnamefont {J.-Y.}\ \bibnamefont {Choi}}, \ and\ \bibinfo
  {author} {\bibfnamefont {B.~H.}\ \bibnamefont {Hong}},\ }\href@noop {}
  {\bibfield  {journal} {\bibinfo  {journal} {Nature}\ }\textbf {\bibinfo
  {volume} {457}},\ \bibinfo {pages} {706} (\bibinfo {year}
  {2009})}\BibitemShut {NoStop}%
\bibitem [{\citenamefont {Lee}\ \emph {et~al.}(2008)\citenamefont {Lee},
  \citenamefont {Wei}, \citenamefont {Kysar},\ and\ \citenamefont
  {Hone}}]{lowD_flexible_2}%
  \BibitemOpen
  \bibfield  {author} {\bibinfo {author} {\bibfnamefont {C.}~\bibnamefont
  {Lee}}, \bibinfo {author} {\bibfnamefont {X.}~\bibnamefont {Wei}}, \bibinfo
  {author} {\bibfnamefont {J.~W.}\ \bibnamefont {Kysar}}, \ and\ \bibinfo
  {author} {\bibfnamefont {J.}~\bibnamefont {Hone}},\ }\href@noop {} {\bibfield
   {journal} {\bibinfo  {journal} {Science}\ }\textbf {\bibinfo {volume}
  {321}},\ \bibinfo {pages} {385} (\bibinfo {year} {2008})}\BibitemShut
  {NoStop}%
\bibitem [{\citenamefont {Castellanos-Gomez}\ \emph {et~al.}(2012)\citenamefont
  {Castellanos-Gomez}, \citenamefont {Poot}, \citenamefont {Steele},
  \citenamefont {van~der Zant}, \citenamefont {Agrait},\ and\ \citenamefont
  {Rubio-Bollinger}}]{lowD_flexible_3}%
  \BibitemOpen
  \bibfield  {author} {\bibinfo {author} {\bibfnamefont {A.}~\bibnamefont
  {Castellanos-Gomez}}, \bibinfo {author} {\bibfnamefont {M.}~\bibnamefont
  {Poot}}, \bibinfo {author} {\bibfnamefont {G.~A.}\ \bibnamefont {Steele}},
  \bibinfo {author} {\bibfnamefont {H.~S.~J.}\ \bibnamefont {van~der Zant}},
  \bibinfo {author} {\bibfnamefont {N.}~\bibnamefont {Agrait}}, \ and\ \bibinfo
  {author} {\bibfnamefont {G.}~\bibnamefont {Rubio-Bollinger}},\ }\href@noop {}
  {\bibfield  {journal} {\bibinfo  {journal} {Nanoscale Research Letters}\
  }\textbf {\bibinfo {volume} {7}},\ \bibinfo {pages} {233} (\bibinfo {year}
  {2012})}\BibitemShut {NoStop}%
\bibitem [{\citenamefont {Wei}\ and\ \citenamefont {Peng}(2014)}]{Wei_2014}%
  \BibitemOpen
  \bibfield  {author} {\bibinfo {author} {\bibfnamefont {Q.}~\bibnamefont
  {Wei}}\ and\ \bibinfo {author} {\bibfnamefont {X.}~\bibnamefont {Peng}},\
  }\href@noop {} {\bibfield  {journal} {\bibinfo  {journal} {Applied Physics
  Letters}\ }\textbf {\bibinfo {volume} {104}},\ \bibinfo {pages} {251915}
  (\bibinfo {year} {2014})}\BibitemShut {NoStop}%
\bibitem [{\citenamefont {Mermin}\ and\ \citenamefont
  {Wagner}(1966)}]{Mermin_Wagner}%
  \BibitemOpen
  \bibfield  {author} {\bibinfo {author} {\bibfnamefont {N.~D.}\ \bibnamefont
  {Mermin}}\ and\ \bibinfo {author} {\bibfnamefont {H.}~\bibnamefont
  {Wagner}},\ }\href@noop {} {\bibfield  {journal} {\bibinfo  {journal}
  {Physical Review Letters}\ }\textbf {\bibinfo {volume} {17}},\ \bibinfo
  {pages} {1133} (\bibinfo {year} {1966})}\BibitemShut {NoStop}%
\bibitem [{\citenamefont {Mehboudi}\ \emph {et~al.}(2016)\citenamefont
  {Mehboudi}, \citenamefont {Dorio}, \citenamefont {Zhu}, \citenamefont
  {van~der Zande}, \citenamefont {Churchill}, \citenamefont {Pacheco-Sanjuan},
  \citenamefont {Harriss}, \citenamefont {Kumar},\ and\ \citenamefont
  {Barraza-Lopez}}]{lowD_P1}%
  \BibitemOpen
  \bibfield  {author} {\bibinfo {author} {\bibfnamefont {M.}~\bibnamefont
  {Mehboudi}}, \bibinfo {author} {\bibfnamefont {A.~M.}\ \bibnamefont {Dorio}},
  \bibinfo {author} {\bibfnamefont {W.}~\bibnamefont {Zhu}}, \bibinfo {author}
  {\bibfnamefont {A.}~\bibnamefont {van~der Zande}}, \bibinfo {author}
  {\bibfnamefont {H.~O.~H.}\ \bibnamefont {Churchill}}, \bibinfo {author}
  {\bibfnamefont {A.~A.}\ \bibnamefont {Pacheco-Sanjuan}}, \bibinfo {author}
  {\bibfnamefont {E.~O.}\ \bibnamefont {Harriss}}, \bibinfo {author}
  {\bibfnamefont {P.}~\bibnamefont {Kumar}}, \ and\ \bibinfo {author}
  {\bibfnamefont {S.}~\bibnamefont {Barraza-Lopez}},\ }\href@noop {} {\bibfield
   {journal} {\bibinfo  {journal} {Nano Letters}\ }\textbf {\bibinfo {volume}
  {16}},\ \bibinfo {pages} {1704} (\bibinfo {year} {2016})}\BibitemShut
  {NoStop}%
\bibitem [{\citenamefont {Wu}\ and\ \citenamefont {Zeng}(2016)}]{lowD_P2}%
  \BibitemOpen
  \bibfield  {author} {\bibinfo {author} {\bibfnamefont {M.}~\bibnamefont
  {Wu}}\ and\ \bibinfo {author} {\bibfnamefont {X.~C.}\ \bibnamefont {Zeng}},\
  }\href@noop {} {\bibfield  {journal} {\bibinfo  {journal} {Nano Letters}\
  }\textbf {\bibinfo {volume} {16}},\ \bibinfo {pages} {3236} (\bibinfo {year}
  {2016})}\BibitemShut {NoStop}%
\bibitem [{\citenamefont {Wang}\ and\ \citenamefont {Qian}(2017)}]{lowD_P3}%
  \BibitemOpen
  \bibfield  {author} {\bibinfo {author} {\bibfnamefont {H.}~\bibnamefont
  {Wang}}\ and\ \bibinfo {author} {\bibfnamefont {X.}~\bibnamefont {Qian}},\
  }\href@noop {} {\bibfield  {journal} {\bibinfo  {journal} {2D Materials}\
  }\textbf {\bibinfo {volume} {4}},\ \bibinfo {pages} {015042} (\bibinfo {year}
  {2017})}\BibitemShut {NoStop}%
\bibitem [{\citenamefont {Seixas}\ \emph {et~al.}(2016)\citenamefont {Seixas},
  \citenamefont {Rodin}, \citenamefont {Carvalho},\ and\ \citenamefont
  {Neto}}]{lowD_SnO}%
  \BibitemOpen
  \bibfield  {author} {\bibinfo {author} {\bibfnamefont {L.}~\bibnamefont
  {Seixas}}, \bibinfo {author} {\bibfnamefont {A.}~\bibnamefont {Rodin}},
  \bibinfo {author} {\bibfnamefont {A.}~\bibnamefont {Carvalho}}, \ and\
  \bibinfo {author} {\bibfnamefont {A.~C.}\ \bibnamefont {Neto}},\ }\href@noop
  {} {\bibfield  {journal} {\bibinfo  {journal} {Physical Review Letters}\
  }\textbf {\bibinfo {volume} {116}},\ \bibinfo {pages} {206803} (\bibinfo
  {year} {2016})}\BibitemShut {NoStop}%
\bibitem [{\citenamefont {Xia}\ \emph {et~al.}(2014)\citenamefont {Xia},
  \citenamefont {Wang},\ and\ \citenamefont {Jia}}]{BP}%
  \BibitemOpen
  \bibfield  {author} {\bibinfo {author} {\bibfnamefont {F.}~\bibnamefont
  {Xia}}, \bibinfo {author} {\bibfnamefont {H.}~\bibnamefont {Wang}}, \ and\
  \bibinfo {author} {\bibfnamefont {Y.}~\bibnamefont {Jia}},\ }\href@noop {}
  {\bibfield  {journal} {\bibinfo  {journal} {Nature Communications}\ }\textbf
  {\bibinfo {volume} {5}},\ \bibinfo {pages} {4458} (\bibinfo {year}
  {2014})}\BibitemShut {NoStop}%
\bibitem [{\citenamefont {Hohenberg}\ and\ \citenamefont {Kohn}(1964)}]{DFT_1}%
  \BibitemOpen
  \bibfield  {author} {\bibinfo {author} {\bibfnamefont {P.}~\bibnamefont
  {Hohenberg}}\ and\ \bibinfo {author} {\bibfnamefont {W.}~\bibnamefont
  {Kohn}},\ }\href@noop {} {\bibfield  {journal} {\bibinfo  {journal} {Physical
  Review}\ }\textbf {\bibinfo {volume} {136}},\ \bibinfo {pages} {864}
  (\bibinfo {year} {1964})}\BibitemShut {NoStop}%
\bibitem [{\citenamefont {Kohn}\ and\ \citenamefont {Sham}(1965)}]{DFT_2}%
  \BibitemOpen
  \bibfield  {author} {\bibinfo {author} {\bibfnamefont {W.}~\bibnamefont
  {Kohn}}\ and\ \bibinfo {author} {\bibfnamefont {L.~J.}\ \bibnamefont
  {Sham}},\ }\href@noop {} {\bibfield  {journal} {\bibinfo  {journal} {Physical
  Review}\ }\textbf {\bibinfo {volume} {140}},\ \bibinfo {pages} {1133}
  (\bibinfo {year} {1965})}\BibitemShut {NoStop}%
\bibitem [{\citenamefont {Kresse}\ and\ \citenamefont
  {Joubert}(1999)}]{vasp_paw}%
  \BibitemOpen
  \bibfield  {author} {\bibinfo {author} {\bibfnamefont {G.}~\bibnamefont
  {Kresse}}\ and\ \bibinfo {author} {\bibfnamefont {D.}~\bibnamefont
  {Joubert}},\ }\href@noop {} {\bibfield  {journal} {\bibinfo  {journal}
  {Physical Review B}\ }\textbf {\bibinfo {volume} {59}},\ \bibinfo {pages}
  {1758} (\bibinfo {year} {1999})}\BibitemShut {NoStop}%
\bibitem [{\citenamefont {Kresse}\ and\ \citenamefont
  {Furthm{\"u}ller}(1996)}]{vasp}%
  \BibitemOpen
  \bibfield  {author} {\bibinfo {author} {\bibfnamefont {G.}~\bibnamefont
  {Kresse}}\ and\ \bibinfo {author} {\bibfnamefont {J.}~\bibnamefont
  {Furthm{\"u}ller}},\ }\href@noop {} {\bibfield  {journal} {\bibinfo
  {journal} {Physical Review B}\ }\textbf {\bibinfo {volume} {54}},\ \bibinfo
  {pages} {11169} (\bibinfo {year} {1996})}\BibitemShut {NoStop}%
\bibitem [{\citenamefont {Perdew}\ \emph {et~al.}(2008)\citenamefont {Perdew},
  \citenamefont {Ruzsinszky}, \citenamefont {Csonka}, \citenamefont {Vydrov},
  \citenamefont {Scuseria}, \citenamefont {Constantin}, \citenamefont {Zhou},\
  and\ \citenamefont {Burke}}]{PBEsol}%
  \BibitemOpen
  \bibfield  {author} {\bibinfo {author} {\bibfnamefont {J.~P.}\ \bibnamefont
  {Perdew}}, \bibinfo {author} {\bibfnamefont {A.}~\bibnamefont {Ruzsinszky}},
  \bibinfo {author} {\bibfnamefont {G.~I.}\ \bibnamefont {Csonka}}, \bibinfo
  {author} {\bibfnamefont {O.~A.}\ \bibnamefont {Vydrov}}, \bibinfo {author}
  {\bibfnamefont {G.~E.}\ \bibnamefont {Scuseria}}, \bibinfo {author}
  {\bibfnamefont {L.~A.}\ \bibnamefont {Constantin}}, \bibinfo {author}
  {\bibfnamefont {X.}~\bibnamefont {Zhou}}, \ and\ \bibinfo {author}
  {\bibfnamefont {K.}~\bibnamefont {Burke}},\ }\href@noop {} {\bibfield
  {journal} {\bibinfo  {journal} {Physical Review Letters}\ }\textbf {\bibinfo
  {volume} {100}},\ \bibinfo {pages} {136406} (\bibinfo {year}
  {2008})}\BibitemShut {NoStop}%
\bibitem [{\citenamefont {Monkhorst}\ and\ \citenamefont {Pack}(1976)}]{mp}%
  \BibitemOpen
  \bibfield  {author} {\bibinfo {author} {\bibfnamefont {H.~J.}\ \bibnamefont
  {Monkhorst}}\ and\ \bibinfo {author} {\bibfnamefont {J.~D.}\ \bibnamefont
  {Pack}},\ }\href@noop {} {\bibfield  {journal} {\bibinfo  {journal} {Physical
  Review B}\ }\textbf {\bibinfo {volume} {13}},\ \bibinfo {pages} {5188}
  (\bibinfo {year} {1976})}\BibitemShut {NoStop}%
\bibitem [{\citenamefont {Togo}\ and\ \citenamefont {Tanaka}(2015)}]{phonopy}%
  \BibitemOpen
  \bibfield  {author} {\bibinfo {author} {\bibfnamefont {A.}~\bibnamefont
  {Togo}}\ and\ \bibinfo {author} {\bibfnamefont {I.}~\bibnamefont {Tanaka}},\
  }\href@noop {} {\bibfield  {journal} {\bibinfo  {journal} {Scripta
  Materialia}\ }\textbf {\bibinfo {volume} {108}},\ \bibinfo {pages} {1}
  (\bibinfo {year} {2015})}\BibitemShut {NoStop}%
\bibitem [{\citenamefont {Mills}\ \emph {et~al.}(1995)\citenamefont {Mills},
  \citenamefont {J{\'{o}}nsson},\ and\ \citenamefont {Schenter}}]{neb}%
  \BibitemOpen
  \bibfield  {author} {\bibinfo {author} {\bibfnamefont {G.}~\bibnamefont
  {Mills}}, \bibinfo {author} {\bibfnamefont {H.}~\bibnamefont
  {J{\'{o}}nsson}}, \ and\ \bibinfo {author} {\bibfnamefont {G.~K.}\
  \bibnamefont {Schenter}},\ }\href@noop {} {\bibfield  {journal} {\bibinfo
  {journal} {Surface Science}\ }\textbf {\bibinfo {volume} {324}},\ \bibinfo
  {pages} {305} (\bibinfo {year} {1995})}\BibitemShut {NoStop}%
\bibitem [{\citenamefont {King-Smith}\ and\ \citenamefont
  {Vanderbilt}(1993)}]{berry}%
  \BibitemOpen
  \bibfield  {author} {\bibinfo {author} {\bibfnamefont {R.~D.}\ \bibnamefont
  {King-Smith}}\ and\ \bibinfo {author} {\bibfnamefont {D.}~\bibnamefont
  {Vanderbilt}},\ }\href@noop {} {\bibfield  {journal} {\bibinfo  {journal}
  {Physical Review B}\ }\textbf {\bibinfo {volume} {47}},\ \bibinfo {pages}
  {1651} (\bibinfo {year} {1993})}\BibitemShut {NoStop}%
\bibitem [{\citenamefont {Page}\ and\ \citenamefont
  {Saxe}(2002)}]{Le_Page_2002}%
  \BibitemOpen
  \bibfield  {author} {\bibinfo {author} {\bibfnamefont {Y.~L.}\ \bibnamefont
  {Page}}\ and\ \bibinfo {author} {\bibfnamefont {P.}~\bibnamefont {Saxe}},\
  }\href@noop {} {\bibfield  {journal} {\bibinfo  {journal} {Physical Review
  B}\ }\textbf {\bibinfo {volume} {65}},\ \bibinfo {pages} {104104} (\bibinfo
  {year} {2002})}\BibitemShut {NoStop}%
\bibitem [{\citenamefont {Elahi}\ \emph {et~al.}(2015)\citenamefont {Elahi},
  \citenamefont {Khaliji}, \citenamefont {Tabatabaei}, \citenamefont
  {Pourfath},\ and\ \citenamefont {Asgari}}]{Elahi_2015}%
  \BibitemOpen
  \bibfield  {author} {\bibinfo {author} {\bibfnamefont {M.}~\bibnamefont
  {Elahi}}, \bibinfo {author} {\bibfnamefont {K.}~\bibnamefont {Khaliji}},
  \bibinfo {author} {\bibfnamefont {S.~M.}\ \bibnamefont {Tabatabaei}},
  \bibinfo {author} {\bibfnamefont {M.}~\bibnamefont {Pourfath}}, \ and\
  \bibinfo {author} {\bibfnamefont {R.}~\bibnamefont {Asgari}},\ }\href@noop {}
  {\bibfield  {journal} {\bibinfo  {journal} {Physical Review B}\ }\textbf
  {\bibinfo {volume} {91}},\ \bibinfo {pages} {115412} (\bibinfo {year}
  {2015})}\BibitemShut {NoStop}%
\bibitem [{\citenamefont {Bosak}\ \emph {et~al.}(2007)\citenamefont {Bosak},
  \citenamefont {Krisch}, \citenamefont {Mohr}, \citenamefont {Maultzsch},\
  and\ \citenamefont {Thomsen}}]{Bosak_2007}%
  \BibitemOpen
  \bibfield  {author} {\bibinfo {author} {\bibfnamefont {A.}~\bibnamefont
  {Bosak}}, \bibinfo {author} {\bibfnamefont {M.}~\bibnamefont {Krisch}},
  \bibinfo {author} {\bibfnamefont {M.}~\bibnamefont {Mohr}}, \bibinfo {author}
  {\bibfnamefont {J.}~\bibnamefont {Maultzsch}}, \ and\ \bibinfo {author}
  {\bibfnamefont {C.}~\bibnamefont {Thomsen}},\ }\href@noop {} {\bibfield
  {journal} {\bibinfo  {journal} {Physical Review B}\ }\textbf {\bibinfo
  {volume} {75}},\ \bibinfo {pages} {153408} (\bibinfo {year}
  {2007})}\BibitemShut {NoStop}%
\bibitem [{\citenamefont {Liu}\ \emph {et~al.}(2014)\citenamefont {Liu},
  \citenamefont {Yan}, \citenamefont {Chen}, \citenamefont {Fan}, \citenamefont
  {Sun}, \citenamefont {Suh}, \citenamefont {Fu}, \citenamefont {Lee},
  \citenamefont {Zhou}, \citenamefont {Tongay}, \citenamefont {Ji},
  \citenamefont {Neaton},\ and\ \citenamefont {Wu}}]{Liu_2014}%
  \BibitemOpen
  \bibfield  {author} {\bibinfo {author} {\bibfnamefont {K.}~\bibnamefont
  {Liu}}, \bibinfo {author} {\bibfnamefont {Q.}~\bibnamefont {Yan}}, \bibinfo
  {author} {\bibfnamefont {M.}~\bibnamefont {Chen}}, \bibinfo {author}
  {\bibfnamefont {W.}~\bibnamefont {Fan}}, \bibinfo {author} {\bibfnamefont
  {Y.}~\bibnamefont {Sun}}, \bibinfo {author} {\bibfnamefont {J.}~\bibnamefont
  {Suh}}, \bibinfo {author} {\bibfnamefont {D.}~\bibnamefont {Fu}}, \bibinfo
  {author} {\bibfnamefont {S.}~\bibnamefont {Lee}}, \bibinfo {author}
  {\bibfnamefont {J.}~\bibnamefont {Zhou}}, \bibinfo {author} {\bibfnamefont
  {S.}~\bibnamefont {Tongay}}, \bibinfo {author} {\bibfnamefont
  {J.}~\bibnamefont {Ji}}, \bibinfo {author} {\bibfnamefont {J.~B.}\
  \bibnamefont {Neaton}}, \ and\ \bibinfo {author} {\bibfnamefont
  {J.}~\bibnamefont {Wu}},\ }\href@noop {} {\bibfield  {journal} {\bibinfo
  {journal} {Nano Letters}\ }\textbf {\bibinfo {volume} {14}},\ \bibinfo
  {pages} {5097} (\bibinfo {year} {2014})}\BibitemShut {NoStop}%
\bibitem [{\citenamefont {Peng}\ and\ \citenamefont {De}(2013)}]{Peng_2013}%
  \BibitemOpen
  \bibfield  {author} {\bibinfo {author} {\bibfnamefont {Q.}~\bibnamefont
  {Peng}}\ and\ \bibinfo {author} {\bibfnamefont {S.}~\bibnamefont {De}},\
  }\href@noop {} {\bibfield  {journal} {\bibinfo  {journal} {Physical Chemistry
  Chemical Physics}\ }\textbf {\bibinfo {volume} {15}},\ \bibinfo {pages}
  {19427} (\bibinfo {year} {2013})}\BibitemShut {NoStop}%
\bibitem [{\citenamefont {Wang}(2003)}]{BFO_1}%
  \BibitemOpen
  \bibfield  {author} {\bibinfo {author} {\bibfnamefont {J.}~\bibnamefont
  {Wang}},\ }\href@noop {} {\bibfield  {journal} {\bibinfo  {journal}
  {Science}\ }\textbf {\bibinfo {volume} {299}},\ \bibinfo {pages} {1719}
  (\bibinfo {year} {2003})}\BibitemShut {NoStop}%
\bibitem [{\citenamefont {Catalan}\ and\ \citenamefont {Scott}(2009)}]{BFO_2}%
  \BibitemOpen
  \bibfield  {author} {\bibinfo {author} {\bibfnamefont {G.}~\bibnamefont
  {Catalan}}\ and\ \bibinfo {author} {\bibfnamefont {J.~F.}\ \bibnamefont
  {Scott}},\ }\href@noop {} {\bibfield  {journal} {\bibinfo  {journal}
  {Advanced Materials}\ }\textbf {\bibinfo {volume} {21}},\ \bibinfo {pages}
  {2463} (\bibinfo {year} {2009})}\BibitemShut {NoStop}%
\bibitem [{\citenamefont {Park}\ \emph {et~al.}(2014)\citenamefont {Park},
  \citenamefont {Le}, \citenamefont {Jeong},\ and\ \citenamefont
  {Lee}}]{BFO_3}%
  \BibitemOpen
  \bibfield  {author} {\bibinfo {author} {\bibfnamefont {J.-G.}\ \bibnamefont
  {Park}}, \bibinfo {author} {\bibfnamefont {M.~D.}\ \bibnamefont {Le}},
  \bibinfo {author} {\bibfnamefont {J.}~\bibnamefont {Jeong}}, \ and\ \bibinfo
  {author} {\bibfnamefont {S.}~\bibnamefont {Lee}},\ }\href@noop {} {\bibfield
  {journal} {\bibinfo  {journal} {Journal of Physics: Condensed Matter}\
  }\textbf {\bibinfo {volume} {26}},\ \bibinfo {pages} {433202} (\bibinfo
  {year} {2014})}\BibitemShut {NoStop}%
\bibitem [{\citenamefont {Duan}\ \emph {et~al.}(2005)\citenamefont {Duan},
  \citenamefont {Sabiryanov}, \citenamefont {Liu}, \citenamefont {Mei},
  \citenamefont {Dowben},\ and\ \citenamefont {Hardy}}]{GdN_1}%
  \BibitemOpen
  \bibfield  {author} {\bibinfo {author} {\bibfnamefont {C.-G.}\ \bibnamefont
  {Duan}}, \bibinfo {author} {\bibfnamefont {R.~F.}\ \bibnamefont
  {Sabiryanov}}, \bibinfo {author} {\bibfnamefont {J.}~\bibnamefont {Liu}},
  \bibinfo {author} {\bibfnamefont {W.~N.}\ \bibnamefont {Mei}}, \bibinfo
  {author} {\bibfnamefont {P.~A.}\ \bibnamefont {Dowben}}, \ and\ \bibinfo
  {author} {\bibfnamefont {J.~R.}\ \bibnamefont {Hardy}},\ }\href@noop {}
  {\bibfield  {journal} {\bibinfo  {journal} {Physical Review Letters}\
  }\textbf {\bibinfo {volume} {94}},\ \bibinfo {pages} {237201} (\bibinfo
  {year} {2005})}\BibitemShut {NoStop}%
\bibitem [{\citenamefont {Liu}\ \emph {et~al.}(2011)\citenamefont {Liu},
  \citenamefont {Ma}, \citenamefont {Zhu},\ and\ \citenamefont {Liu}}]{GdN_2}%
  \BibitemOpen
  \bibfield  {author} {\bibinfo {author} {\bibfnamefont {H.~M.}\ \bibnamefont
  {Liu}}, \bibinfo {author} {\bibfnamefont {C.~Y.}\ \bibnamefont {Ma}},
  \bibinfo {author} {\bibfnamefont {C.}~\bibnamefont {Zhu}}, \ and\ \bibinfo
  {author} {\bibfnamefont {J.-M.}\ \bibnamefont {Liu}},\ }\href@noop {}
  {\bibfield  {journal} {\bibinfo  {journal} {Journal of Physics: Condensed
  Matter}\ }\textbf {\bibinfo {volume} {23}},\ \bibinfo {pages} {245901}
  (\bibinfo {year} {2011})}\BibitemShut {NoStop}%
\end{thebibliography}
\end{document}